\documentclass{article}
 \usepackage{graphicx}

\def\beq{\begin{equation}}
\def\eeq{\end{equation}}
\def\bea{\begin{eqnarray}}
\def\eea{\end{eqnarray}}
 \def\r{{\bf r}}
 
\def\p{{\bf p}}
 
 \def\half{\frac{1}{2}}

 \def\half{{\frac{1}{2}}}
 
\begin{document}
 
\title{\bf  An extended Vlasov equation\\
  with pairing correlations}
\author{ V. I. Abrosimov$^{\rm a}$, D. M. Brink$^{\rm b}$,A.Dellafiore $^{\rm c}$,
F.Matera$^{\rm c}$\\
\it $^{\rm a}$ Institute for Nuclear Research,  Kiev, Ukraine\\
 $^{\rm b}$ \it Oxford University, Oxford, U.K.\\ 
$^{\rm c}$ \it Istituto Nazionale di Fisica Nucleare, Sezione di Firenze\\
\it  and Dipartimento di Fisica, Universit\`a degli Studi di Firenze,\\
\it via Sansone 1, I-50019 Sesto F.no (Firenze), Italy}
\date{}
\maketitle
\begin{abstract}

A simplified version of the Wigner--transformed time--dependent Hartree--Fock--Bogoliubov equations, leading to a solvable model for  finite systems of fermions with pairing correlations, is introduced. In this model, pairing correlations result in a coupling of the Vlasov--type equation for the normal phase--space density with that for  the imaginary part of the anomalous density. The effect of pairing correlations on the linear response of the system is studied for a finite one--dimensional system and an explicit expression for the correlated propagator is given.
\end{abstract}

\section{ Introduction}

The collisionless Boltzmann equation, or Vlasov equation,  is a useful tool for studying quantum systems in the semiclassical limit (\cite{rs}, p. 553).

It is well known that the Vlasov kinetic equation can be derived from the time--dependent Hartree--Fock
equation

\beq
\label{tdhf}
i\hbar\partial_t \rho=[h,\rho]\,
\eeq
by taking the Wigner transform of the density matrix $\rho$, however,
a straightforward application of this formalism does not give immediately the Vlasov equation:  the additional approximation of neglecting the momentum dependence of the Wigner--transformed Hartree--Fock potential is required. 

A similar derivation for the time--dependent Hartree--Fock--Bogoliubov (TDHFB) equations would be desirable for applications both in nuclear physics \cite{dk} and in the physics of trapped fermion droplets \cite{us}, however the Wigner--transformed TDHFB equations are a rather complicated set of coupled differential equations \cite{us} for the normal and anomalous densities $\rho(\r,\p,t)$ and $\kappa(\r,\p,t)$
($\rho$ is real, while $\kappa$ is complex) and it is not obvious which approximations should be introduced in order to obtain the equivalent of the Vlasov equation for superfluid systems.

We have studied such an approximation that has the advantage of leading to simple solvable equations,
 even if it has the problem of violating particle-number conservation.  We shall see that this problem can be solved within the model. Here we discuss this model  for a one-dimensional system of spin-saturated fermions. Like for the normal Vlasov equation, understanding the simpler one--dimensional problem is an important preliminary step for studying  three--dimensional spherical systems \cite{bdd}.  By comparing the response of the correlated system to an external perturbation with that of a normal system described by the ordinary Vlasov equation, we can learn about the  effects of pairing correlations on the system dynamics. 

A fundamental role in determining the system eigenfrequencies is played by the boundary conditions imposed on the fluctuations of the density, here we make the comparison between correlated and uncorrelated systems by employing the simple fixed--surface boundary conditions of \cite{bdd}, but fancier (moving--surface) boundary conditions might also be considered \cite{ads} (see also \cite{abr05}).

\section{Static limit}

Bengtsson and Schuck \cite{bs}  have studied the semiclassical limit of the static HFB equations (see also \cite{rs}, p. 550). By following their method, we find  the self-consistent equilibrium solutions\footnote{Note that our sign of $\kappa_0$  differs from that given in Eq. (9) of \cite{bs}, this means that our $\kappa$ corresponds to $-\kappa$ of Refs. \cite{bs}, \cite{us} and \cite{rs}.}  
\bea
\label{roz}
&&\rho_{0}(\r,\p)=\frac{1}{2}{\Big(}1-\frac{h_0(\r,\p)-\mu}{E(\r,\p)}{\Big)}\,\\
&&\kappa_0(\r,\p)=-\frac{\Delta_0(\r,\p)}{2E(\r,\p)}\,,
\label{kappaz}
\eea

where
\begin{itemize}

\item{$h_0(\r,\p)$ is the self--consistent Hartree--Fock equilibrium hamiltonian,}
\item{ the chemical potential $\mu$   is determined by the number of particles:
\beq
\label{norma}
A=\frac{g}{(2\pi\hbar)^3}\int d\r d\p\rho_0(\r,\p)
\eeq
($g$ is the number of fermions in a phase--space cell, $g=4$ for nucleons),}
\item{$E(\r,\p)$ is the quasiparticle energy, defined as
\beq
\label{qpe}
E(\r,\p)=\sqrt{\Delta_0^2(\r,\p)+(h_0(\r,\p)-\mu)^2}\,,
\eeq
}
\item{$\Delta_0(\r,\p)$ is the equilibrium pairing field, which is related to $\kappa_0(\r,\p)$  also by the self--consistency relation (\cite{rs}, p. 550)\footnote{In view of the previous remark on the sign of $\kappa$, here the sign of $v(|\p-{\bf k}|)$ must be opposite to that used in \cite{rs}}
\beq
\Delta_0(\r,\p)=\frac{1}{(2\pi\hbar)^3}\int d{\bf k}v(|\p-{\bf k}|)\kappa_0(\r,{\bf k})\,.
\eeq
}
\end{itemize}
We take a more phenomenological approach and replace the  pairing field $\Delta_0(\r,\p)$ with the phenomenological pairing gap, in heavy nuclei $\Delta\approx 1{\rm MeV}$. In this case the static solutions become
\bea
\rho_0(\epsilon)&=&\frac{1}{2}{\Big (}1-\frac{\epsilon-\mu}{E(\epsilon)}{\Big )}\,,\\
\label{kazero}
\kappa_0(\epsilon)&=&-\frac{\Delta}{2 E(\epsilon)}\,,\\
E(\epsilon)&=&\sqrt{\Delta^2+(\epsilon-\mu)^2}\,,
\eea
with $\epsilon=h_0(\r,\p)$ the particle energy.
Note that
\beq
\label{kapparo}
\kappa_0(\epsilon)=\frac{E^2(\epsilon)}{\Delta}\frac{d\rho_0(\epsilon)}{d\epsilon}\,.
\eeq

In Fig.1 the functions $\rho_0(\epsilon)$ and $\kappa_0(\epsilon)$ are plotted for typical values of parameters $\mu=33 {\rm MeV}$ and $\Delta=1{\rm MeV}$.
\begin{figure}[h]
\vspace{.2in}
\centerline {
\includegraphics[width=3in]{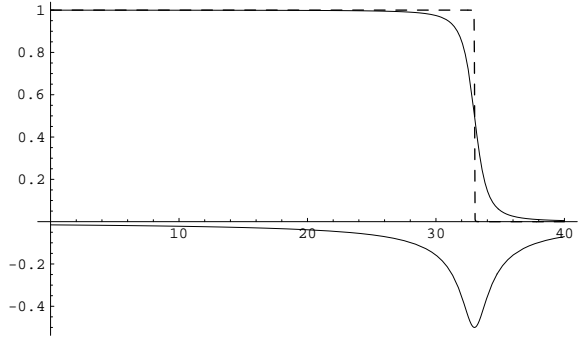}
}
\caption{Normal (solid, upper) and anomalous (solid, lower) densities as a function of particle energy, expressed in MeV. The dashed curve shows the normal distribution for $\Delta=0.01\,{\rm MeV}$.}
\vspace{.2in}
\end{figure}

\section{Linear response}
We want to study the linear response of a system that, in equilibrium conditions, is described by the equations above. We assume that we have a finite, spin--saturated system of fermions at zero temperature with pairing correlations that, at time $t=0$ is subject to a weak perturbing external force  generated by a driving field of the kind $\delta V^{ext}(\r,t)=\beta\delta(t)Q(\r)$ (the parameter $\beta$ determines the intensity of the external force, the $\delta$--function its time dependence and $Q(\r)$ its spatial distribution). We  want to determine the fluctuations of the density of the system at point $\r$ as a function of time for $t>0$. This will be done in linear approximation, that is, we neglect terms of second order or higher in the fluctuations.  Thus we have:
\bea
h(\r,\p,t)&=&h_0(\r,\p)+\delta h(\r,\p,t)=\epsilon+\delta V^{ext}(\r,t)\,,\\
\rho(\r,\p,t)&=&\rho_0(\epsilon)+\delta\rho(\r,\p,t)\,,\\
\kappa(\r,\p,t)&=&\kappa_0(\epsilon)+\delta\kappa_r(\r,\p,t)+i\delta\kappa_i(\r,\p,t)
\eea
and, from the TDHFB equations, the following coupled equations for $\delta\rho(\r,\p,t)$ and $\delta\kappa_i(\r,\p,t)$ can be obtained:
\bea
\label{norm}
\partial_t\delta\rho(\r,\p,t)&=&\{h,\rho\}- 2\frac{\Delta}{\hbar}\delta\kappa_i(\r,\p,t)\\
\label{anom}
\partial_t\delta\kappa_i(\r,\p,t)&=&2\frac{E^2(\epsilon)}{\hbar\Delta}{\Big(}\half[\delta\rho(\r,\p,t)+\delta\rho(\r,-\p,t)]\nonumber\\
& &-\frac{d\rho_0}{d\epsilon}\delta h(\r,\p,t){\Big)}\,.
\eea
These two coupled equations can be interpreted as an extension of the ordinary Vlasov equation of normal systems to systems with pairing. In the limit $\Delta\to 0$, the first equation gives the usual Vlasov equation, while the second equation can be ignored. When  $\Delta\ne 0$ the two equations are coupled. Here we consider only the approximation $\delta h(\r,\p,t)=\delta V^{ext}(\r,t)$, which corresponds to the single--particle approximation of the quantum approach, taking into account also the mean--field fluctuations induced by the external field leads to collective effects (see \cite{bdd}).

Taking the Fourier transform in time of Eqs. (\ref{norm}, \ref{anom}), gives
\bea
-i\omega\delta \rho(\r,\p,\omega)&=&\{h, \rho\}-2\frac{\Delta}{\hbar}\delta\kappa_i(\r,\p,\omega)\,,\\
-i\omega\delta\kappa_i(\r,\p,\omega))&=&2\frac{E^2(\epsilon)}{\hbar\Delta}{\Big (}\half[\delta \rho(\r,\p,\omega)+\delta \rho(\r,-\p,\omega)]\nonumber\\
&-&\frac{d\rho_0}{d\epsilon}\delta h(\r,\p,\omega){\Big )}\,,\eea
or, for $\omega\neq 0$,
\newpage
\bea
\label{val}
-i\omega \delta \rho({\bf r},{\bf p},\omega) +\{\delta \rho, h_{0}\}
&=& -i\omega d^2\half[\delta \rho(\r,\p,\omega)+\delta \rho(\r,-\p,\omega)]\nonumber\\
&-&\frac{d\rho_0}{d\epsilon}{\big(}\{h_0,\delta h\}-i\omega d^2\delta h{\big )}\,,
\eea
with 
\beq
d^{2}={\Big (}\frac{\Omega(\epsilon)}{\omega}{\Big )}^2
\eeq
and 
\beq
\Omega(\epsilon)=2\frac{E(\epsilon)}{\hbar}\,.
\eeq
 In Eq. (\ref{val}) we have used the relation $\{\rho_0,\delta h\}=\frac{d\rho_0}{d\epsilon}\{h_o,\delta h\}$. The equation corresponding to Eq.(\ref{val}) for normal systems can be formally obtained from this equation by letting $d^2\to 0$. (cf. Eq. (2.12) of \cite{bdd}). Thus the effect of pairing in the present approximation is only that of adding the terms containing $d^2$ in Eq. (\ref{val}).

\section{One--dimensional system}
For one--dimensional systems it is convenient to make the change of variables $(x,p_x)\to(x,\epsilon)$, with $\epsilon=\frac{p_x^2}{2m}+V_0(x)$.  Then Eq. (\ref{val}) gives the following system of coupled differential equations:

\bea
\label{aa}
&&\frac{\partial f^{+}}{\partial x}-\frac{i\omega}{v(\epsilon,x)}f^{+}=B^+(x)- 
\frac{i\omega}{v(\epsilon,x)}d^{2}\frac{1}{2}[f^{+}+f^{-}]\\
\label{bb}
&&\frac{\partial f^{-}}{\partial x}+\frac{i\omega}{v(\epsilon,x)}f^{-}=B^-(x)+ 
\frac{i\omega}{v(\epsilon,x)}d^{2}\frac{1}{2}[f^{+}+f^{-}]\,,
\eea
 where
 \beq
 f^\pm=\delta \rho(x,\pm\sqrt{2m[\epsilon-V_0(x)]},\omega)\,
 \eeq

and the inhomogeneous term 
\beq
B^\pm(x)=B(x)\pm d^2 C(x)
\eeq
 contains both the inhomogeneous term of the normal Vlasov equation
\beq
B(x)=\frac{d \rho_0}{d\epsilon}{\Big (}\beta\frac{d Q(x)}{dx}{\Big )}
\eeq
and an additional term proportional to the anomalous density (cf. Eq. (\ref{kapparo}))
\beq
C(x)=\frac{i\omega}{v(\epsilon,x)}\frac{d\rho_0}{d\epsilon}{\Big (}\beta Q(x){\Big)}\,.
\eeq
We have neglected the contribution of mean--field fluctuations to the inhomogeneous terms $B(x)$ and $C(x)$, so we limit our discussion to the zero--order approximation of \cite{bdd}. In this case the coupled system (\ref{aa}, \ref{bb}) can be solved exactly and its solution allows us to write an expression of the correlated propagator analogous to that obtained in \cite{bdd} for the uncorrelated propagator.

The uncorrelated propagator obtained in \cite{bdd} was
\bea
\label{unco}
&&D^0(x,x',\omega)=\frac{g}{2\pi\hbar}\\
&&2\int d\epsilon\frac{d\rho_0}{d\epsilon}\sum_{n}\frac{-2n\omega_0}{T}\frac{\cos[n\omega_0\tau(x)]}{v(\epsilon,x)}\frac{1}{\omega-n\omega_o+i\varepsilon}\frac{\cos[ n\omega_0\tau(x')]}{v(\epsilon,x')}\,,\nonumber
\eea
 while the correlated propagator given by the solution of the system (\ref{aa}, \ref{bb}) is
\bea
\label{corr}
&&\tilde D^0(x,x',\omega)=\frac{g}{2\pi\hbar}\\
&&2\int d\epsilon \frac{d\rho_0}{d\epsilon}\sum_{n}\frac{-2\tilde\omega_n}{T}\frac{\cos[n\omega_0\tau(x)]}{v(\epsilon,x)}\frac{1}{\omega-\tilde\omega_n+i\varepsilon}\frac{\cos [n\omega_0\tau(x')]}{v(\epsilon,x')}\,.\nonumber
\eea

By comparing  the  expressions of the two propagators, we can  immediately appreciate  some of the changes induced by the pairing correlations:
\begin{itemize}
\item{The equilibrium distribution is changed from the familiar Fermi--gas step function (dashed curve in Fig. 1) to the smoother curve also shown in Fig. 1, so the function $\frac{d\rho_0}{d\epsilon}$, which is a $\delta$--function in the uncorrelated case, becomes also smoother and this smears  the response.}

\item{The position of the propagator's poles are determined by the eigenfrequencies of the system. For the uncorrelated propagator they are $\omega_n=n\omega_0$, where $n$ is an arbitrary integer and $\omega_0=2\pi/T$  is determined  by the system size (for given $\epsilon$), while for the correlated propagator, the eigenfrequencies become
\bea
\label{eifr}
\tilde\omega_n&=&n\omega_0\sqrt{1+{\Big(}\frac{\Omega(\epsilon)}{n\omega_0}{\Big)^2}}\qquad \:\:{\rm for}\; n\neq0\,,\\
\tilde\omega_n&=&\pm\Omega(\epsilon)\qquad\qquad\qquad\qquad{\rm for}\;n=0\,.
\eea
}
\end{itemize}

A closer analysis of the correlated response function
\beq
\tilde S^0(\hbar\omega)=-\frac{1}{\pi}{\rm Im}\int  dx\int dx' Q(x)\tilde D^0(x,x',\hbar\omega) Q(x')
\eeq
reveals  other interesting properties of the correlated system:

\begin{itemize}
\item{the correlated response function displays a gap of $\approx2\Delta$ at low excitation energy;}
\item{the lack of self--consistency due to the constant--$\Delta$ approximation generates some spurious strength in the correlated response.}
\end{itemize}

There are two sources of spurious strength: one is related to the mode $n=0$ and its possible contribution to particle--number non conservation. However, even if the contribution of this mode is omitted from the propagator (\ref{corr}), the resulting strength function gives an energy--weighted sum rule
\beq
\tilde M_1=\int_0^\infty d\hbar\omega\,\hbar\omega\tilde S^0(\hbar\omega)\approx M_1[1+14.66(\frac{\Delta}{\hbar\omega_0})^2]
\eeq
which differs from the uncorrelated sum rule
\beq
M_1=\frac{2}{3}\hbar^2\frac{AL^2}{m}\,.
\eeq
What is even more disturbing is the fact that the enhancement factor  depends on the size of the system.
Both problems can be solved by introducing a modified propagator
\bea
\label{phys}
&&\tilde D^0_{phys}(x,x',\omega)=\frac{g}{2\pi\hbar}\\
&&2\int d\epsilon \frac{d\rho_0}{d\epsilon}\sum_{n}(\frac{n\omega_0}{\tilde\omega_n})^2\frac{-2\tilde\omega_n}{T}\frac{\cos[n\omega_0\tau(x)]}{v(\epsilon,x)}\frac{1}{\omega-\tilde\omega_n+i\varepsilon}\frac{\cos [n\omega_0\tau(x')]}{v(\epsilon,x')}\,.\nonumber
\eea

This modified propagator contains no contribution from the mode $n=0$ and this ensures particle--number conservation, since
\beq
\delta A(\omega)=\int dx dp_x\delta\rho(x,px,\omega)=\int dx dx'\tilde D^0_{phys}(x,x',\omega) Q(x')
\eeq
vanishes for any external field $Q(x')$  if only modes with $n\neq0$ are included in the propagator. Moreover, it can be easily checked that the strength function given by this propagator satisfies the same sum rule as that given by the uncorrelated propagator.
% \newpage

The following two figures show examples of correlated and uncorrelated response function for two systems of different size. The comparison is made by assuming hat the systems have the same value of $\mu$, hence the number of particles is slightly different.

Figure 2 shows a comparison of the first peak in the correlated and uncorrelated strength functions for the external field $Q(x)=x^2$ and an equilibrium mean field $V_0(x)$ of the square--well type. The response functions have been divided by a factor $\frac{g}{2\pi\hbar}L^4$, where $L$ is the size of the square--well mean field, which has been chosen so that $\hbar\omega_0(\mu)=10\, {\rm MeV}$.

Figure 3 shows another example of response to the same external field $Q(x)=x^2$, but in this case the size of the potential well has been chosen so that $\hbar\omega_0(\mu)=1\,{\rm MeV}$. All other parameters are unchanged.  In this case $\hbar\tilde\omega_{n=1}(\mu)\approx 2.23\,{\rm MeV}$ and the main peak appearing in this figure (solid curve) corresponds to the mode $n=1$, the smaller bumps at larger energy are due to the modes with larger values of $n$. Note the gap of $\sim\,2\Delta$ in the low--energy region, the small tail extending into the gap is due to the finite value of $\varepsilon$ used in the calculation.

\newpage

\begin{figure}[h]
\vspace{.2in}
\centerline {
\includegraphics[width=2.5in]{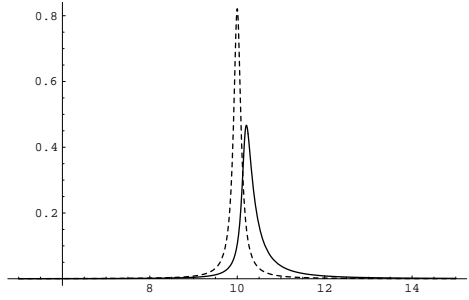}
}
\caption{Uncorrelated and correlated response functions (renormalized), as a function of excitation energy $\hbar\omega$ expressed in MeV. The peaks shown here correspond to the mode $n=1$ of the two propagators (\ref{unco}) and (\ref{corr}).The peak on the left should be a $\delta$-function and it represents the response of a normal system given by the propagator (\ref{unco}); for numerical reasons we have used a finite value of $\varepsilon=0.1\,{\rm MeV}$, thus inducing an artificial smearing of the response. The peak on the right is obtained from the propagator (\ref{phys}) with the same value of $\varepsilon$. Its larger width is an effect of the pairing correlations.There are other similar peaks around $\hbar\omega=20,\, 30\,\ldots\,{\rm MeV}$,  but their strength decreases rapidly with increasing $n$.}
\vspace{.2in}
\end{figure}
\newpage

\begin{figure}[h]
\vspace{.2in}
\centerline {
\includegraphics[width=3in]{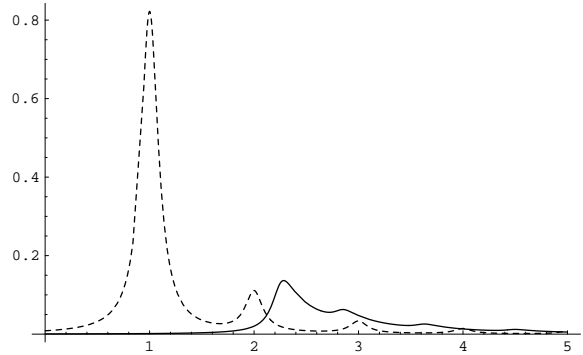}
}
\vspace{.2in}
\caption{Same as Fig.2, but with $\hbar\omega_0(\mu)=1\,{\rm MeV}$ and including modes up to $n=10$. The main peak of the uncorrelated strength function (dashed) at $\hbar\omega=1\,{\rm MeV}$ is pushed to higher energy by the pairing correlations and a gap of about 2$\Delta$ is created at low energy. }
\end{figure}

\end{document}